\documentclass[%
reprint,
amsmath,amssymb,
prb,
]{revtex4-2}

\usepackage{graphicx}
\usepackage{dcolumn}
\usepackage{bm}

\begin{document}


\title{Restructuring in bimetallic core--shell nanoparticles: Real time observation}

\author{Nobutomo Nakamura}
\email{nobutomo@me.es.osaka-u.ac.jp}
\author{Koji Matsuura}%
\author{Akio Ishii}%
\affiliation{%
Graduate School of Engineering Science, Osaka University, Toyonaka, Osaka 560-8531, Japan
}%

\author{Hirotsugu Ogi}
\affiliation{
Graduate School of Engineering, Osaka University, Suita, Osaka 565-0871, Japan
}%


\begin{abstract}
The formation process of core--shell bimetallic nanoparticles synthesized by sputtering onto a substrate is observed in real time using an originally developed acoustic technique. The technique enables us to evaluate the structural change of nanoparticles at room temperature without contacting the nanoparticles or substrate. In the experiments, the sputtering of metal A followed by metal B tended to form B-shell/A-core nanoparticles. However, in Pd--Au alloy system, notable restructuring occurred during synthesis, resulting in the formation of A-shell/B-core nanoparticles. The formation process is analyzed using the molecular dynamics simulation, revealing that this restructuring occurs on a short timescale, and high diffusivity of Au plays an important role. 
\end{abstract}

\maketitle


\section{Introduction}
Core--shell nanoparticles exhibit unique properties that differ from those of single metal nanoparticles. Studies of the relationship between the properties and internal structures of core--shell nanoparticles have been widely conducted with the aim of synthesizing core--shell nanoparticles with arbitrary desired properties \cite{Chaudhuri_ChemRev_2012, Gawande_ChemSocRev_2015,Liao_Small_2015}.

In bimetallic alloys, theoretically, surface segregation occurs depending on several factors such as the surface energy, cohesive energy, and atomic size of each metal, and preferred core--shell structures have been investigated based on these factors \cite{Ferrando_ChemRev_2008,Wang_JACS_2009}. 
However, in experiments, structure of nanoparticles often changes depending on the synthesis methods, and core--shell nanoparticles whose structure is opposite that predicted based on such factors are formed. 
For example, in AuPd alloys, Au is expected to segregate to the surface \cite{Wang_JACS_2009,Rousset_PRB_1996,Marchal_JPCC_2013}. When AuPd nanoparticles are synthesized on a substrate through deposition, the segregation of Au on the surface (formation of Au-shell/Pd-core structure) occurs \cite{Luo_JPCB_2005, Sitja_JPCC_2019} even if Pd is deposited on Au nanoparticles \cite{Haire_SurfSci_2011}. Conversely, when nanoparticles are synthesized through the chemical deposition of Pd on Au nanoparticles \cite{Hu_ChemPhysLett_2005,Rajoua_JPCC_2015} and one-step aqueous synthesis \cite{Lee_JACS_2009}, Pd-shell/Au-core structure is formed. 
This variance of the experimental results from theoretical predictions would be due to the effects of factors such as adsorption of molecules on the surface by exposure to air or liquid during a given experiment, which is not considered in simulations and theoretical calculations. 
Thus, despite the published discussions of preferred core--shell structures, theoretical predictions and experimental results do not always coincide.
Under the circumstances, if structural changes during synthesis are observed in real time, factors affecting the formation process will be identified more clearly, leading to a deeper understanding of the process.

Core--shell nanoparticle characterization is typically performed via transmission electron microscopy, x-ray photoelectron spectroscopy, x-ray diffraction, ultraviolet--visible spectroscopy, and other techniques \cite{Rajoua_JPCC_2015,Guo_JPCC_2011,Bao_JPCC_2007,Alayoglu_NatMater_2008,Langlois_NanoLett_2015}, which are post-characterization methods. For core--shell nanoparticles synthesized through dewetting using pulse-laser heating, \textit{in situ} characterization has been performed using dynamic transmission electron microscope (DTEM) \cite{McKeown_AdvanMater_2015}. However, real-time characterizations during physical and chemical depositions at room temperature have not been performed, although they are representative synthesis methods. For solving this problem, we propose a methodology for the dynamic structural evaluation of core--shell nanoparticles based on their shapes.

When metals are deposited on a substrate, nucleation and nucleus growth result in the formation of spherical-cap-shaped nanoparticles. Core--shell nanoparticles can be obtained by depositing the core metal followed by the shell metal. If the surface energy of the shell metal is higher than that of the core metal, the nanoparticles become spherical to reduce the surface area; in other words, the contact angle increases. On the other hands, if interdiffusion occurs in the nanoparticles and the core--shell structure is not formed, the nanoparticle shape will be different from that of the core--shell structure. Thus, by measuring the nanoparticle shapes, we can evaluate their microscopic structures. However, the direct and dynamic measurement of the shape is difficult. To solve this problem, we propose the use of electrical resistance measurements. When nanoparticles are dispersed on a substrate and their shapes are changed, the gap distances between the nanoparticles change, and the electrical resistance changes with the gap distance \cite{Nishiura_ThinSolidFilms_1974}. Therefore, the change in shape can be evaluated based on the change in resistance. We apply it here to core--shell nanoparticles.

To evaluate nanoparticle shape change via the resistance change, it is necessary to synthesize isolated nanoparticles with extremely narrow gaps, e.g., less than 1 nm. The conduction between isolated nanoparticles is governed by the tunneling current; the narrower the gap, the lower the resistance. Metal nanoparticles do not deform considerably, and the resistance change via shape change is expected to be very small. If the gaps are extremely narrow, the nanoparticles are expected to contact each other with only slight shape changes. This change in contact state switches the mechanism of electrical conduction from tunneling current to bulk conduction, resulting in a large decrease in resistance. This decrease makes it possible to detect small shape changes via resistance change measurements. The remaining difficult is the synthesis of nanoparticles with such narrow gaps. To solve this problem, we use resistive spectroscopy. This method, originally developed to monitor morphological changes in nanoparticles during deposition, has been used to synthesize nanoparticles with extremely narrow gaps on a substrate \cite{Nakamura_JAP_2015, Nakamura_APL_2017, Nakamura_RSI_2021}. This method measures the electrical resistance between nanoparticles using the resonant vibration of a piezoelectric resonator, and the gap distance is evaluated based on the resonance results. This method can thus be used not only to fabricate narrow-gap nanoparticles but also to measure shape change in core--shell nanoparticles via resistance change.

\section{Experimental}

Here we investigate Pd--Ag, Pd--Pt, and Pd--Au nanoparticles; Pd--based core--shell nanoparticles are well known for their efficient catalytic properties in several reactions \cite{Chaudhuri_ChemRev_2012,Tedsree_NatNanotech_2011}. Each type of core--shell nanoparticle was synthesized by sequential sputtering of two metals on a substrate. Al$_{2}$O$_{3}$(0001) plate of 0.43-mm thickness and cover glass of 0.15-mm thickness were used as substrates. The background pressure was less than $3.8\times 10^{-4}$ Pa, and Ar pressure during sputtering was 0.8 Pa. Two cathodes were attached to the sputtering chamber; the two metals were deposited in turn by opening and closing the shutters of each cathode.

When metal is deposited on a substrate, isolated nanoparticles are formed at first, and as they grow, a continuous film is formed. During this morphological transition, the gaps between nanoparticles become narrower, and the electrical resistance of the substrate surface decreases monotonically. In our experiments, a rectangular parallelepiped piezoelectric resonator measuring $2.5\times1.7\times0.2 \mathrm{mm}^3$, made of single crystal LiNbO$_{3}$, was placed beneath the substrate, and the resonant spectrum around the 1.86-MHz resonant mode was measured repeatedly using line antennas and a network analyzer \cite{Nakamura_APL_2017}. When the resonator vibrates at a resonant frequency, an electric field is excited around it, and the field reaches the substrate surface. The field causes electrons to move between nanoparticles, thereby increasing the energy loss of the resonant vibration depending on the electrical resistance. Therefore, by measuring the energy loss, the change in electrical resistance (gap distance) can be evaluated \cite{Nakamura_APL_2017}.

Figure 1(a) shows typical resonant spectra obtained at different times during the deposition of Pd on Al$_{2}$O$_{3}$ (see Supplemental Material 1 \cite{SupplementalMaterial}). The resonant peak is temporarily broadened during the deposition. The time variation of the full width at half maximum (FWHM) of the resonant peak in Fig. 1(a) shows a clear peak around 260 s in Fig. 1(b). In our previous work \cite{Nakamura_APL_2017,Nakamura_APL_2019}, we confirmed that the FWHM peak appears when the nanoparticles contact one another. Therefore, by interrupting deposition before the FWHM peak appears, we synthesized nanoparticles with extremely narrow gaps. 

\begin{figure}
\includegraphics{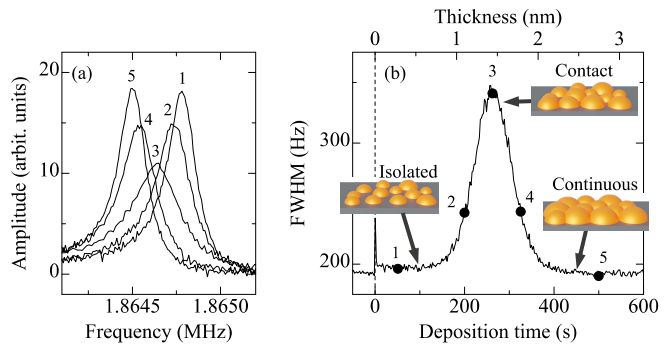}
\caption{(a) Representative resonant spectra measured during deposition of Pd on Al$_{2}$O$_{3}$ and (b) the corresponding FWHM change. Individual points 1--5 in (b) denote the timings at which the corresponding resonant spectra in (a) were obtained. Schematic images of nanoparticles on a substrate are also shown in (b).}
\end{figure}

In the following experiments, metal A is deposited first to fabricate core nanoparticles. Metal B is then deposited to fabricate core--shell nanoparticles. These nanoparticles are hereafter described as B/A nanoparticles. 
During the deposition of metal A, the gaps between nanoparticles become monotonically narrower, and the FWHM peak appears. However, if the nanoparticle shapes are altered by depositing metal B on top of metal A, the gap distance does not change monotonically, and the shape of the FWHM peak changes accordingly. Therefore, nanoparticle shape change can be evaluated based on the time evolution of FWHM.

\section{Results and discussion}
\subsection{Monitoring of structural change using resistive spectroscopy}
Figure 2(a) shows FWHM changes during deposition on an Al$_{2}$O$_{3}$ substrate for the Pd--Ag system. Pd and Ag were deposited at deposition rates of 0.02 and 0.014 nm/s, respectively. When Ag or Pd was singly and continuously deposited, a single FWHM peak appeared around 540 or 65 s, respectively. In contrast, when Pd was deposited after Ag, two FWHM peaks appeared. When the deposition metal was changed after the first FWHM peak (Pd/Ag (i)), the FWHM transiently increased and then decreased again (see Supplemental Material 2 \cite{SupplementalMaterial}). When the deposition metal was changed while the FWHM was increasing (Pd/Ag (ii)), the FWHM transiently decreased and then showed an FWHM peak (see Supplemental Material 3 \cite{SupplementalMaterial}). These behaviors can be explained as follows. The surface energy of Pd is higher than that of Ag by 0.76 J/m$^{2}$ at 0 K \cite{Tyson_SurfSci_1977}. Therefore, when Pd is deposited on Ag nanoparticles, the surface energy of the nanoparticles increases and the nanoparticles tend to become more spherical. The electrical resistance therefore increases as the gaps between the nanoparticles become wider or as the nanoparticles that were in contact with each other detach. When the shape change is complete, the nanoparticles again grow in the plane parallel to the substrate, and the electrical resistance decreases as the gaps become narrower and the nanoparticles come into contact.
\begin{figure}
\includegraphics{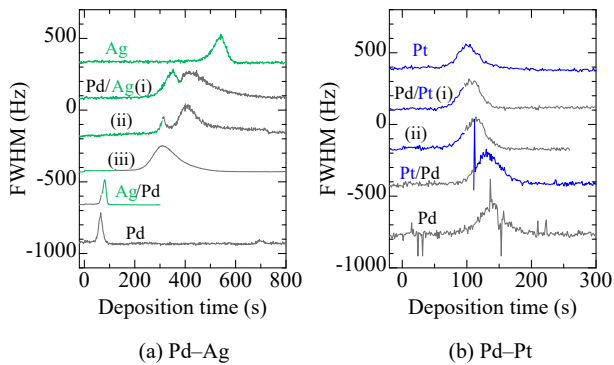}
\caption{FWHM changes during deposition of (a) Pd--Ag system and (b) Pd--Pt system. Data are shifted in the vertical direction.}
\end{figure}

When the deposition metal was changed before the FWHM started to increase (Pd/Ag (iii)), only one FWHM peak was observed. In this scenario, changing the deposition metal changed the shapes of the nanoparticles. However, because resistive spectroscopy is not sensitive to shape changes in nanoparticles with wide gaps, shape change was not detected, and a single FWHM peak appeared. When Ag was deposited after Pd was deposited (Ag/Pd), the FWHM curve became slightly more peaked. This finding is attributed to the low surface energy of Ag, which causes the nanoparticles to elongate in the in-plane direction, accelerating the narrowing of gaps.

Figure 2(b) shows the FWHM time variation of the Pd--Pt system on the Al$_{2}$O$_{3}$ substrate; each of Pd and Pt was deposited at a rate of 0.01 nm/s. Considering that the surface energy of Pt is higher than that of Pd by 0.49 J/m$^{2}$ \cite{Tyson_SurfSci_1977}, the Pt/Pd nanoparticles were expected to produce two FWHM peaks. However, in all experiments (Pd/Pt and Pt/Pd), a single FWHM peak was observed, and the peak duration was almost constant. A possible cause is the immiscibility of the PdPt alloy; the PdAg alloy is a miscible system \cite{Massalski_Birany_1990}, whereas the PdPt alloy is an immiscible system \cite{Okamoto_JPhaseEq_1991}. Considering that phase separation occurs in immiscible systems, Pt and Pd nanoparticles may grow independently. In that case, the change in the timing of the FWHM peak should be interpreted as the average change over two scenarios, i.e., when Pt or Pd is deposited. 

For the Pd--Au system, it was expected that the Pd/Au nanoparticles would produce two FWHM peaks, because it is a miscible system, and the surface energy of Au is lower than that of Pd by 0.50 J/m$^{2}$ \cite{Tyson_SurfSci_1977}. However, as seen in Fig. 3(a), unexpected behaviors were observed. Pd and Au were deposited at deposition rates of 0.0055 and 0.0052 nm/s, respectively. When the deposition of Pd was initiated just before the FWHM reached its maximum (Pd/Au (i)), the subsequent change in FWHM became very gradual, and the FWHM peak was greatly broadened (see Supplemental Material 4 \cite{SupplementalMaterial}). This very slow decrease in the FWHM indicates that in-plane nanoparticle growth was suppressed and that the nanoparticles grew in the out-of-plane direction with additional deposition. This suppression of in-plane growth was also observed when the deposition metal was changed before the FWHM began increasing (Pd/Au (ii)). When either Pd or Au was singly deposited continuously, the FWHM peak appeared at around 260 or 630 s, respectively, but for Pd/Au (ii) nanoparticles, the FWHM peak appeared at around 1760 s. These results indicate that when Pd is deposited on Au nanoparticles, in-plane growth is strongly suppressed. The same behaviors were observed on a glass substrate as shown in Fig. 3(b), indicating that the suppression is intrinsic to Pd/Au nanoparticles, and insensitive to the substrate material. Considering that the surface energy of Pd is higher than that of Au, the explanation for this result can be qualitatively similar to that of the Pd--Ag system results. However, in-plane growth is significantly suppressed in the Pd--Au system, and the formation dynamics of Pd/Au nanoparticles seem to differ from those of Pd/Ag nanoparticles. When Au was deposited after Pd (Au/Pd), the FWHM peak tended to become sharp, especially on the glass substrate as shown in Fig. 3(b). This behavior can be explained by the lower surface energy of Au as in the case of Ag/Pd nanoparticles.
\begin{figure}
\includegraphics{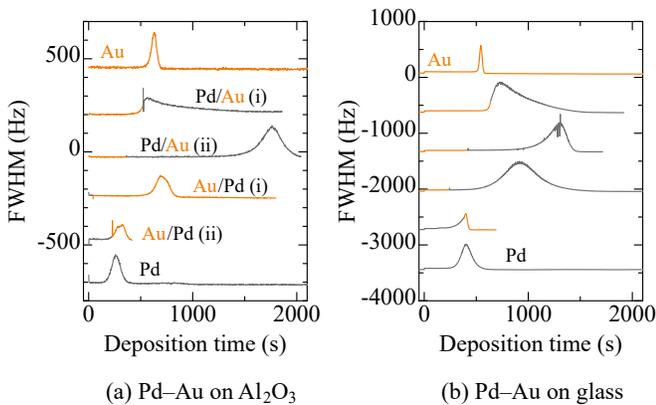}
\caption{FWHM changes during deposition of Pd--Au on (a) Al$_{2}$O$_{3}$ and (b) glass. Data are shifted in the vertical direction.}
\end{figure}

In our previous work on deposition of single metal on quartz substrate \cite{Nakamura_JAP_2015}, the FWHM gradually changed with time after interrupting the deposition around the FWHM peak due to the shape change of the nanoparticles. Similar behaviors were observed in the deposition of Au and Pd on glass substrate (see Supplemental Material 5 \cite{SupplementalMaterial}). These results imply that the FWHM change observed after changing the deposition metal might be caused by the shape change of the core nanoparticles, not by the change in the surface energy by shell-metal deposition.  However, the FWHM change after interrupting the deposition of single metal was slow and gradual compared to that after changing the deposition metal.  Therefore, we consider that the FWHM change observed in the above experiments is mainly caused by the change in the surface energy by the shell-metal deposition. 

\subsection{Molecular dynamics simulation}

To investigate changes in atomic structure during deposition, we performed molecular dynamics (MD) simulation using B/A cuboidal face-centered-cubic wafer models. In Fig. 4(a), we show a schematic of our wafer model. Our wafer model includes A (bottom) and B (top) atomic wafers, mimicking the first and second sputtered metallic layers, respectively. 
The surroundings of wafers were set as a vacuum, and a wall was set at $\rm Z=0$, mimicking the substrate in our physical experiment. 
Wall--atom interaction for both A and B atoms were described as a Lennard--Jones potential, mimicking the interaction between the substrate and the sputtered atoms. 
Embedded atom method potentials \cite{Shan_PRB_2009,Shan_PRB_2014,Hale_ModelSimuMaterSciEng_2013,Zhou_PRB_2004} were used as interaction potentials between metals. Wall--atom interactions were set to be weaker than atom--atom interactions.
LAMMPS~\cite{Plimpton_JComputPhys_1995} code was used for the MD simulation itself. 
The initial atomic structure of the Pd/Ag wafer model is shown in Fig. 4(b) as an example. 
After initial structural relaxation, we implemented 3--ns NVT ensemble simulation at temperatures of 300, 400, and 600 K for each wafer model. The simulation time of 3~ns may seem short. However, in the study using DTEM \cite{McKeown_AdvanMater_2015}, structural change in nanosecond timescale is observed, and the trend of structural evolution will be observed within this simulation timeframe. Details of the simulation conditions are described in Supplemental Material 5\cite{SupplementalMaterial}.
\begin{figure}
\includegraphics[width=76mm]{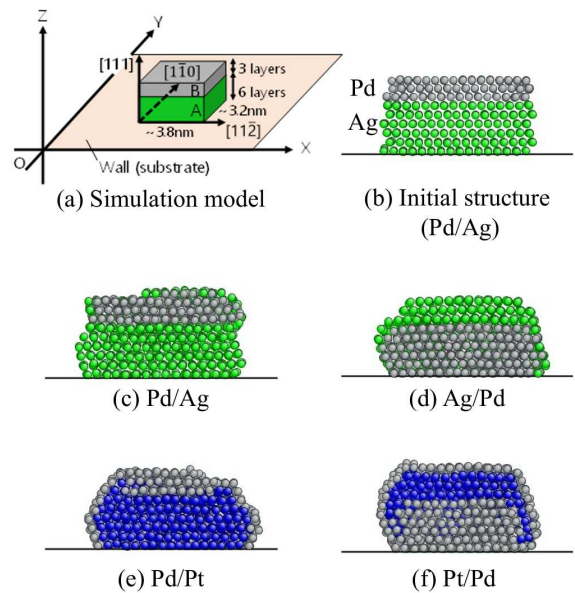}
\caption{(a) Schematic of a wafer model for MD simulation. (b) Initial structure of the Pd/Ag model. Cross sectional views (X--Z plane) of the (c) Pd/Ag, (d) Ag/Pd, (e) Pd/Pt, and (f) Pt/Pd models after the ensemble simulation at 600 K for 3 ns. (gray: Pd, green: Ag, and blue: Pt)}
\end{figure}
\begin{figure}
\includegraphics[width=70mm]{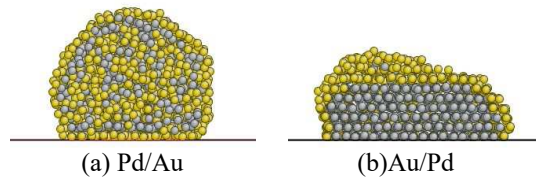}
\caption{Cross-sectional view (X--Z plane) of the (a) Pd/Au and (b) Au/Pd models at 400 K. (gray: Pd and yellow: Au)}
\end{figure}

Figures 4(c) and 4(d) show the cross-sections of the Pd/Ag and Ag/Pd models, respectively, after 3--ns simulation at 600 K (Data for 300 K and 400 K simulations are shown in Supplemental Material 5\cite{SupplementalMaterial}). In the Ag/Pd model, Ag atoms diffuse on the surface of the Pd layer, making the nanoparticles wider and accelerating in-plane growth. Conversely, in the Pd/Ag model, Ag atoms barely diffuse to cover the Pd layer, and Pd-shell/Ag-core structure is maintained. Consequently, the nanoparticle hardly grows in the in-plane direction. These results are in agreement with our experimental results. 

In the Pd--Pt system (Fig. 4(e), Fig. 4(f)), Pd diffused onto the Pt surface, and the surfaces of both Pd/Pt and Pt/Pd nanoparticles were covered with Pd. There was no significant difference in shape between the two models. These results are consistent with our experimental results, in which the trends in FWHM during deposition were almost the same for Pd/Pt and Pt/Pd nanoparticles (Data for 300 K and 400 K simulations are shown in Supplemental Material 5\cite{SupplementalMaterial}).

In the Pd--Au system, as seen in Fig. 5, interdiffusion occurred at 400 and 600 K for Pd/Au nanoparticles, as well as at 600 K for Au/Pd nanoparticles (see Supplemental Material 5 \cite{SupplementalMaterial}). In the models, spherical particles were formed. Based on these results, interdiffusion occurs in Pd--Au system even at low temperatures, and the formation of spherical nanoparticles and segregation of Au are accompanied with it. Such restructuring reduces the growth rate of nanoparticles in the in-plane direction, thereby promoting growth in the out-of-plane direction; closure of the gap distances is therefore slow, resulting in a broad FWHM peak. Although 400 K is higher than the room temperature conditions of the experiments, the temperature of the metal on the substrate is expected to be higher than room temperature during sputtering. 

Considering that the trend (slope) of the time evolution of FWHM changes immediately after the sputtered metal is changed (second data from the top in Fig. 3(b)), the time scales of interdiffusion and subsequent restructuring are comparable to or shorter than the time step of the FWHM measurement, typically a few seconds. Assuming that the time step $t$ is 5 s and diffusion length $x$ is 4 nm, the typical thickness of Au nanoparticles at the FWHM peak, the interdiffusion coefficient $D$ of $1.6 \times 10^{-14}$ cm$^2$/s is obtained as $D=x^2/2t$. This value is much higher than the diffusion coefficient of bulk PdAu at room temperature, $\sim10^{-28}$ cm$^2$/s, which is extrapolated from the high-temperature values \cite{Neukam_Gal_1970}. Regarding the Pd/Au film, the interdiffusion coefficients of the order of $10^{-15}-10^{-13}$ cm$^2$/s at 350 $^\circ$C \cite{Murakami_JAP_1976} and $\sim10^{-13}$ cm$^2$/s at room temperature (extrapolated from the high-temperature values \cite{Boiko_FizMetlMet_1968}) are reported. The present value is close to the latter one, but actual value may be higher, because the time step of the FWHM measurement is not short enough to monitor rapid structural change. In a previous study \cite{Haire_SurfSci_2011}, the segregation of Au in Pd/Au nanoparticles on oxide surfaces at room temperature was observed using medium energy ion scattering, post-characterization, and it was attributed to the high diffusivity of Au. Extrapolating the surface diffusion coefficients of Au, Ag, and Pd on Al$_2$O$_3$ at high temperatures \cite{Erdelyi_JAP_1996,Imre_SurfSci_1999,Beszeda_ApplPhysA_2005} to 300 K, the coefficient of Au becomes much higher than that of Ag and Pd (Au: $2.1\times 10^{-24}$, Ag: $3.0\times 10^{-37}$, Pd: $2.7\times 10^{-40}$ m$^2$/s). It implies that high diffusivity of Au atoms on substrate surface also contributes to the shape change and segregation.

\section{Conclusions}
In summary, we developed the method for the dynamic structural evaluation of core--shell nanoparticles, and it was revealed that formation process of core--shell nanoparticles differs in different bimetallic nanoparticles. For Pd--Ag nanoparticles, Ag atoms tend to diffuse on Pd to segregate on nanoparticle surface, but they cannot diffuse sufficiently at around room temperature. Consequently, the nanoparticle shape changes depending on the surface energy of the metal sputtered on core nanoparticles. 
For Pd--Pt nanoparticles, notable differences were not observed in the formation processes of Pd/Pt and Pt/Pd nanoparticles; immiscibility was considered as a cause. 
For Pd--Au nanoparticles, when Au is sputtered on Pd nanoparticles, the surface diffusion of Au causes the elongation in the plane direction. In contrast, when Pd is sputtered on Au nanoparticles, interdiffusion occurs, and segregation and shape change are accompanied with it. As described in this paper, factors dominantly affecting the formation process can be identified using the developed method, and the method will contribute the development of core--shell nanoparticles with arbitrary structures.

\begin{acknowledgments}
This research was partially supported by JSPS KAKENHI Grant Numbers 18H01883 and 20K21145.
\end{acknowledgments}

\bibliography{PdAu_nanoparticle_PRB.bib}

\end{document}